# Transient Luminous Events in the lower part of the Atmosphere originated in the Peripheral Regions of a Thunderstorm

A.Chilingarian, G.Hovsepyan, T.Karapetyan, B.Sargsyan, E.Svechnikova

Yerevan Physics Institute, Alikhanyan Brothers 2, Yerevan, Armenia, 0036

**Abstract.** We present and discuss Transient Luminous Events (TLEs) in the lower atmosphere, observed during large disturbances of the near-surface electric fields (NSEF) and coinciding with large enhancements of the particle fluxes (thunderstorm ground enhancements – TGEs). In spite of large distances from the strongest electric field region the maximum energy of TGE particles on 22 and 25 May 2018 reaches 50 MeV. Thus, the accelerating electric field reaches 2.0 keV/cm even far from the zone of the strong lightning activity on the periphery of the storm. The light glows appearing at the same time in the skies can be due to the local charge rearrangement generating a small illuminating discharge without initiating the lightning flash. These unusual luminous phenomena are suggested that an electrical discharge much weaker than a lightning flash, could only partially neutralize the charge aloft, and hence, only partially lower the corresponding potential difference, allowing the electron accelerator to operate and send particle fluxes in the direction to the earth's surface. Simultaneously, these types of discharges initiate light glows in the thunderous atmosphere inside and below thunderclouds.
Correlation analysis of enhanced particle fluxes demonstrates that the radiation-emitting region in the thundercloud can be larger than 1 km.

1. Introduction

Thunderstorm ground enhancements (TGEs, [1,2]) observed mostly in high mountain areas by a variety of particle detectors are large impulsive enhancements of the electrons and gamma ray (rarely also neutrons [3,4]), fluxes lasting from tens of seconds to tens of minutes and sometimes exceeding the cosmic ray background hundreds of times [5]. These enhancements imply that very specific conditions of the atmospheric electric field are established inside the cloud, leading to the multiplication and acceleration of electrons [6]. The essence of these conditions is the development of an electrical dipole (in the lower part of the thundercloud), which accelerates electrons and decelerates positrons and muons in the direction of the ground [7]. If the atmospheric electric field exceeds the critical value, specific to air density (height in the atmosphere), electrons runaway and produces relativistic runaway electron avalanches (RREA) [8-10]. The scenarios of the RREA process initiating large TGEs are discussed, see Figure 1 in [11]. Comparison of measured TGEs and simulations with GEANT4 and CORSIKA codes allow one to outline the plausible vertical profile of the electric field necessary for starting a runaway process.

Research in the past two decades has identified a surprising variety of "Transient Luminous Events" (TLEs), which accompany thunderstorms. Following the trend to look for the possible optical counterparts of thunderstorms, we establish on Aragats 24/7 monitoring of the skies by panoramic cameras as part of the multisensory research in the high-energy atmospheric physics. In recent years, we register a wide range of light structures, like large blobs in the center of the field of view (FoW) of the camera, near-vertical luminous filaments, narrow jets reaching the ground, multiple small blobs covering almost the entire FoW of the panoramic camera. The predominant color of the light glows is blue-violet. In order to place the optical observations in an electric field structure and particle fluxes context, we examine TGEs and disturbances of the near-surface electric field (NSEF) simultaneously with the light glows. The sky monitoring system allows to store 1-minute time-series across all hours of the day, with occasionally appearing lightning flashes, light glows, birds, and meteors. We found that the optical flashes are not isolated events, but happened during active thunderstorms, on the periphery of the strong lightning activity, usually accompanied by TGEs and special types of disturbances of NSEF. Peripheral storm regions do not produce lightning flashes, nonetheless produce enigmatic light glows.

Possibly, discharges, from which the glows originated, begin as lightning leaders inside the thundercloud, which were not able to collect enough charge to develop -IC or -CG lightning, and end up as a partial discharge illuminating the skies only. These, "intermediate" types of discharges, between lightning flashes and corona discharges, occurred when the net potential drop isn't enough to initiate a "real" flash.

Measurements of TGEs at the periphery of the thunderstorm, as well, give clues to understanding the horizontal extension of the electric field, that supports RREA development in large areas. The size of the particle emitting region in a thundercloud still remains not well researched. Measurements with multiple dosimeters installed at nuclear power plants in a coastal area of the Japanese sea made it possible to follow the source of the gamma ray flux moving with an ambient wind flow [12]. At Nor Amberd research station, located on slopes of Mt. Aragats at 2000 m height, the size of the particle emitting region was estimated Using the muon stopping effect [13, 14]. Estimates from both studies locate particle emitting regions within 1 km. However, in the recent radar-based gamma glow (TGE) study along the coast of the Japanese sea, it was observed that all TGEs were accompanied by the graupel fall, indicating the low location of the lower positively charged region [15]. A strong radar echo due to the high reflectivity of hydrometeors, indicates that the vertical and horizontal extent of the strong accelerating electric field was larger than 2 km. In another observation of the gamma glow in Japan the flux enhancements were initiated and terminated exactly at the same time at a distance of 1.35 km [16]. Thus, the previously estimated values of particle emitting region size within 1 km seem to be highly underestimated.

We use facilities of the Aragats research station operated by cosmic ray division (CRD) of the Yerevan Physics Institute, i.e., networks of particle detectors, NSEF sensors, lightning locators, and panoramic cameras, to relate the light glows to the special conditions of the electrified atmosphere, and - to estimate the size of the particle emitting region using a large collection of TGEs registered on Aragats during last decade.

2. **Instrumentation**

In Fig. 1 we show the location of the 2 main networks used in this study. In Fig. 2a we depict the GOOGLE map with located EFM-100 electric mills produced by BOLTEC firm widely used in the atmospheric physics research. EFM-100 electric mills are measuring NSEF with a frequency of 20 HZ and send measurements via WIFI to online computers, and then to the CRD's MySQL database. Usually, we use 1-s averaged time series of the NSEF for the multivariate visualization and correlation analysis. We locate electric field sensors on masts due to deep snow in the winter months. The mast heights on which the electric field sensors are located vary from 3 to 13 m above the ground. EFM-100 sensor estimates as well the distance to the lightning flash within 33 km from the sensor with an accuracy of ≈1.5 km. The comparisons with WWLN estimates of the distance to lightning from the station show a good agreement within the accuracies of both lightning location systems [17].

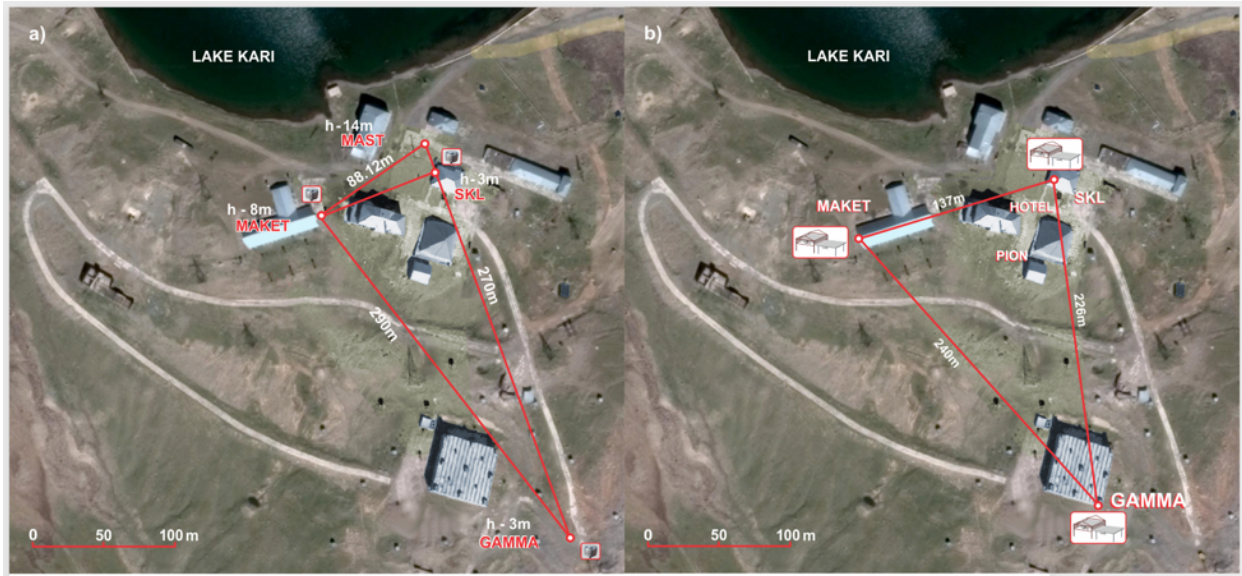

**Figure 1. a) EFM-100 electric mills network, near 4 sites the hights of masts are indicated; b) STAND1 particle detectors network location.**

In Fig. 1b we show the network of STAND1 particle detectors placed in the vertices of a triangle sides of which are equal to 133 m, 226 m, and 240 m. The "STAND1" detector is comprised of three layers of 1-cm-thick, 1-m$^2$ sensitive area scintillators stacked vertically and one 3-cm thick plastic scintillator of the same type stands apart; see Fig. 2. The light from the scintillator is reradiated into the long-wavelength region of the spectrum by the spectrum-shifter fibers and passed to the photomultiplier (PMT FEU-115M). The maximum luminescence is emitted at the 420-nm wavelength, with a luminescence time of about 2.3 ns [18]. The STAND1 detector is tuned by changing the high voltage applied to the PMT and by setting the thresholds for the shaper-discriminator. The discrimination level is chosen to guarantee both high efficiency of signal detection and maximal suppression of photomultiplier noise.

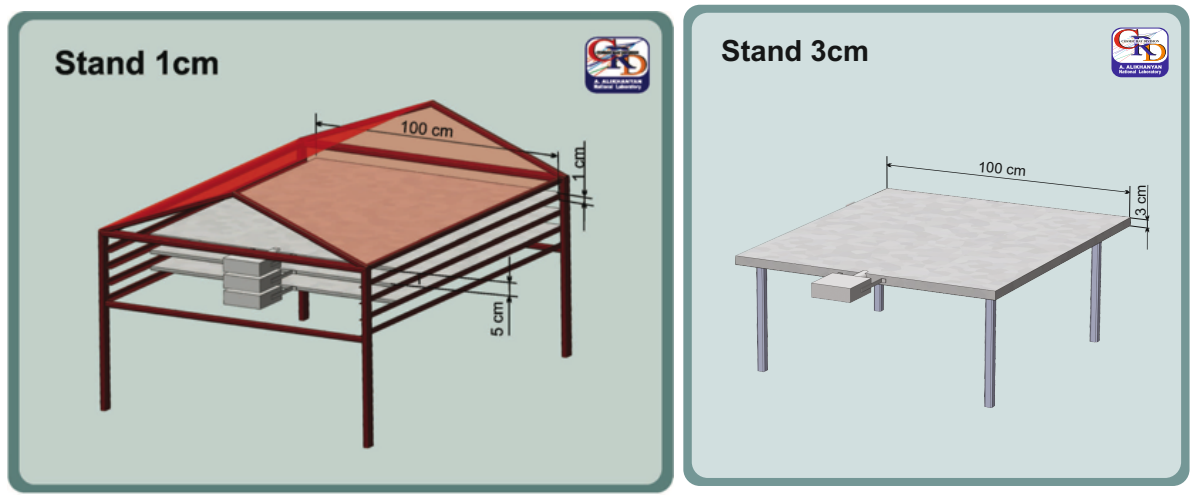

**Figure 2. STAND1 detector consisting of three layers of 1-cm- thick scintillators and stand-alone 3-cm thick plastic scintillator of the same type.**

From simulations and from calibration experiments we estimate the efficiency of the STAND1 scintillators for charged particles as ≈95%. Consequently, the probability to miss a particle is ≈5%. Using the energy spectrum recovered by the NaI spectrometers we estimate with the GEANT4 code the probability of registering a gamma ray by the upper, middle, and lower layers of the STAND1 detector to be ≈2%, ≈2.5%, and ≈2.8%, respectively. Using these efficiencies, and the energy spectrum of the TGE gamma rays measured by NaI spectrometers, we estimate the number of electrons registered in each of the 3 layers. Afterward, proceeding from the energy threshold of each scintillator and 3 count rates we can roughly estimate the electron energy spectrum.

DAQ system used for the STAND1 network is the National Instrument's MyRio board [19]. It includes eight analog inputs, four analog outputs, 32 digital I/O lines, programmable FPGA, and a dual-core ARM Cortex-A9 processor (a high-performance processor implementing the full richness of the widely supported ARMv7-A architecture). With reconfigurable FPGA technology, we perform high-speed signal processing, high-speed control, inline signal processing, and custom timing and triggering. Eight digital inputs of 3 myRIO boards are used for feeding signals from 3STAND1 detectors (4 channels each), proportional counters of Aragats neutron monitor (ArNM), and NaI crystal-based spectrometers. The myRIO pulse counting system can provide registration of very short time series (down to 1 ms) that enables the investigation of the dynamic of TGE development and its relation to the lightning initiation (50 ms time series are stored currently).

The structure of the atmosphere during TGE events was modeled using a numerical mesoscale model: the Weather Research and Forecasting Model (Advanced Research WRF (WRF-ARW), v. 4.1.2) [20, 21]. The strategy of two nested domains is applied, with the center at the observation point (40.4715°, 44.1815°). The outer domain with dimensions of 2,700 × 1,800 km$^2$ (the discretization step is 3 km), the inner domain with dimensions of 90 × 90 km$^2$ (the discretization step is 1 km) covers the Aragats research station. The vertical coordinate in the inner domain has an irregular grid of 41 levels, the discretization step, which was changing from 50 m near the ground to 600 m at the height of 20 km. WRF model includes various parameterizations describing physical processes on a sub-grid scale, including microphysical processes. The set of parameterizations providing the most reliable modeling for clouds producing TGE observed at the Aragats Station is based on the recommendations for fine meshes (MP_PHYSICS = 8, RA_LW_PHYSICS = 4, RA_SW_PHYSICS = 4, RADT = 10, SF_SFCLAY_PHYSICS = 1, SF_SURFACE_PHYSICS = 2, BL_PBL_PHYSICS = 2) and a verification procedure using the results of near-surface measurement, satellite data and data of meteorological radar located 25 km from the Aragats station [22, 23].

3. Thundercloud extension, electric field strength and particle fluxes

The thundercloud coverage of Armenia and, especially, its presence above Aragats research station we estimate by the map of lightning locations with Boltek's Storm Tracker (lightning detection system [24]), powered by the software from Astrogenics. Storm tracker defines four types of lightning types (CG-, CG+ cloud-to ground negative and positive, IC -, IC+ intracloud positive and negative) in radii up to 480 km around the location of its antenna. By the examining time-slices of the lightning activity we determine from which direction the storm is coming, and finally, putting on the map all lightning occurrences we can see if the storm's active zone misses the station (Fig. 3).

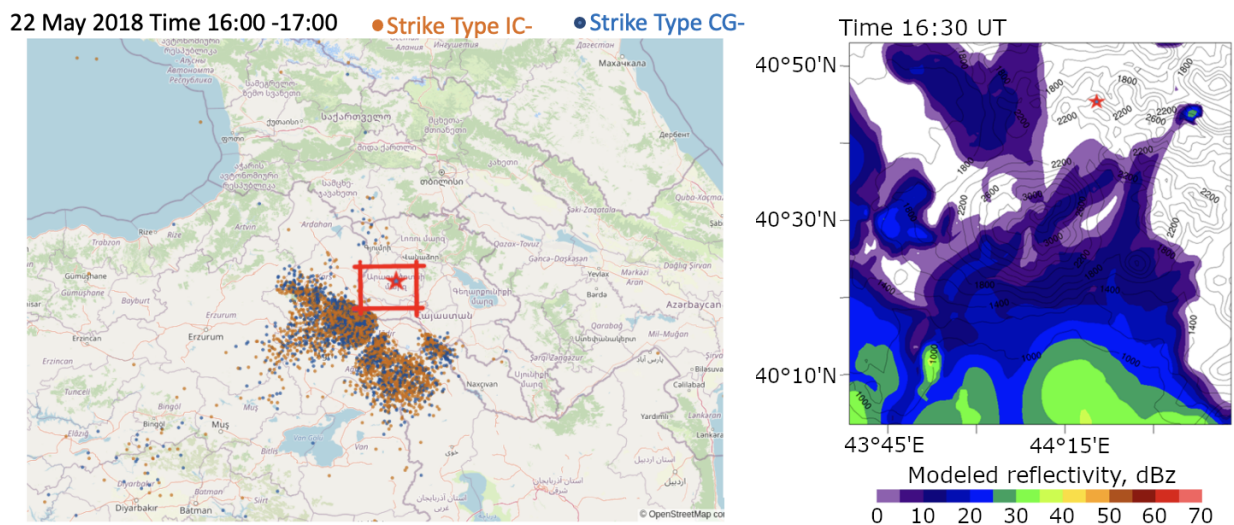

**Figure 3. Left: Projection of the thundercloud obtained by registering the lightning active zone with Boltek's storm tracker, 22 May 2018 15:16 – 21:53. The lightning type symbols and storm duration are shown in the upper panel, the Aragats research station location is shown by the red star. Right: The maximum of the modeled with WRF radar reflectivity in the air column, 22 May 2018 16:30. Red squares mark the same region in both pictures.**

In Fig.3 by mapping all flashes of a huge storm coming from the south-west we show that the storm passed aside the Aragats station on May 22, and, therefore, TGE wasn't terminated by the lightning flash as usual, and continued for 18 minutes, as we can see in Fig. 4e.

In Fig. 4e we show 1-s time series of 3 modules of the STAND1 network (upper 1-cm thick scintillators). All three count rates are precisely correlated, the correlation between the MAKET-SKL detectors (Fig. 4a, distance 88m, correlation coefficient – 0.81) are the same as the correlation between the MAKET-GAMMA detectors (Fig. 4c, distance 240m, correlation coefficient – 0.82). There is no significant shift in the times of the maximum flux measured by pairs of detectors, as it is shown by the plot of the "delayed" correlations in Figs. 4b) and 4d). A delayed correlation plot is obtained by shifting time series one relative to another by 1 to 200 seconds. If there is a delay in particle arrivals, the maximum correlation will be shifted, and not peaked on the 0 value. Thus, all 3 detectors show precisely coinciding time series of TGE count rates.

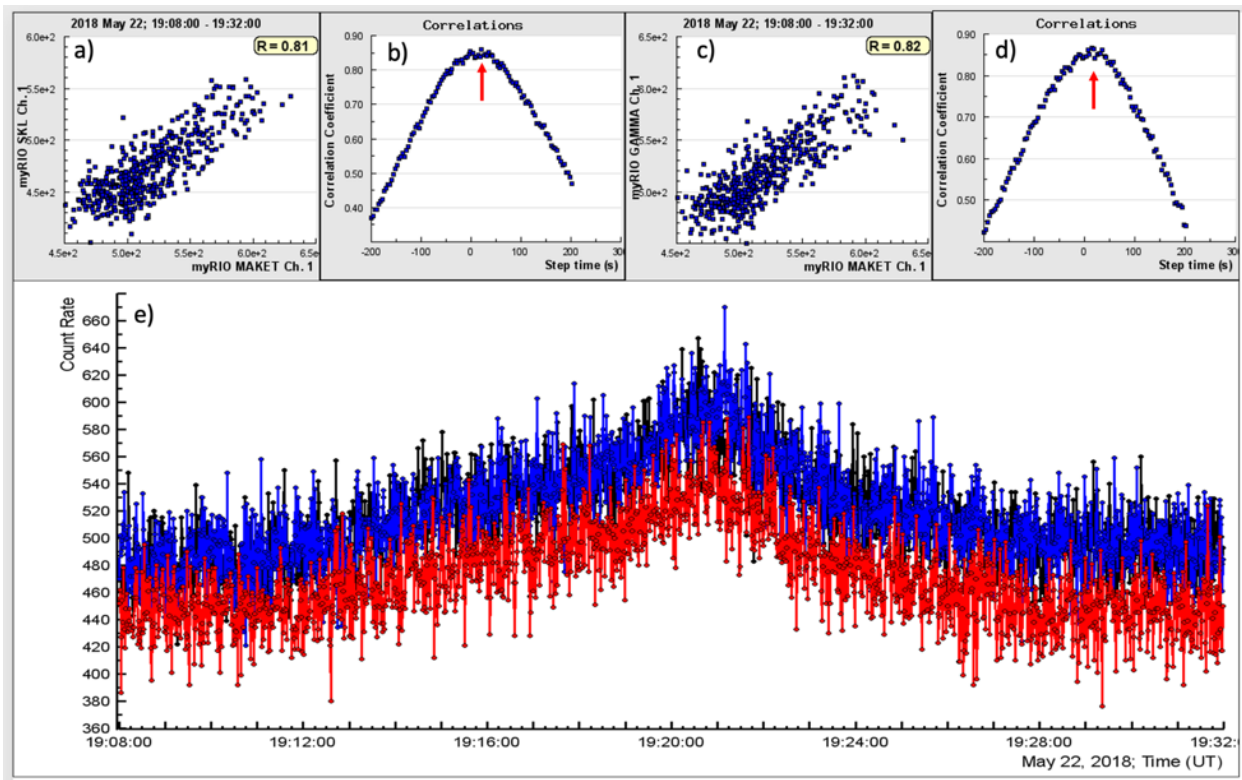

**Figure 4. a) and b) scatter plot and delayed correlations plot (with a shift from -200 to 200 seconds) of count rates of STAND1 detector's upper 1-cm thick scintillators (MAKET and SKL detectors); c) and d) the same for the MAKET and GAMMA detectors; e) 1-sec time series of all 3 units of STAND1 network: MAKET (black), GAMMa (blue), and SKL (red).**

In Fig.5 we show the meteorological conditions observed during the TGE event. The outside temperature was 1.8C, and relative humidity also was stable -95% during TGE; the cloud base was as low as 87.5 m. inset a), nearest lightning flash - on 10 km from the detectors, inset b). Cloud base height was estimated by calculating the spread between the air temperature and dew point according to the well-known approximate equation (see, details in WEB calculators):
**H(m) ≈ (Air temperature at surface {°C} – dew point temperature {°C}) x 125**
It was raining during the TGE (red line).

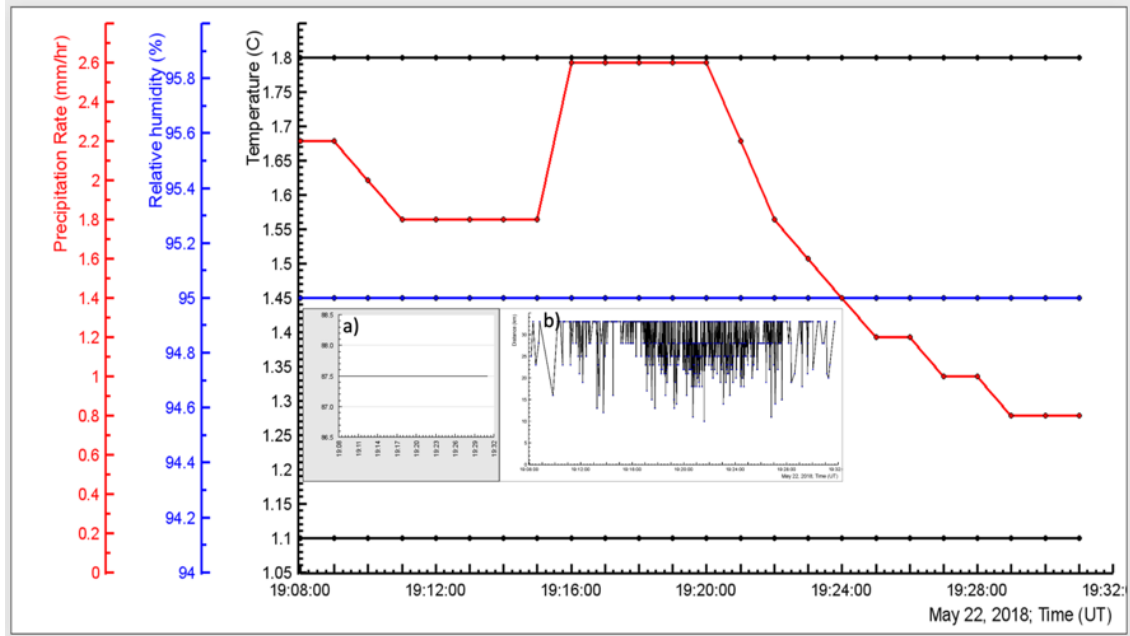

**Figure 5. 1-min time series of outside temperature and dew point (black lines); relative humidity (blue line); and rainfall (red line). In the inset a) we show the estimated distance to cloud base, in the inset b) – distances to lightning flashes occurred during TGE.**

In Fig.6 we show the 1-s time series of the NSEF, measured by the electric mill located on the roof of the MAKET experimental hall. NSEF was in the deep negative domain during TGE, with 3 outbursts of the NSEF with an amplitude of ≈10kV/m occurring during 3 minutes (19:19-19:21) coinciding with TGE maximum flux (see Fig. 4e). These outbursts coincide also with the maximum of the differential energy spectrum of the particle flux registered by the network of NaI spectrometers, see Fig. 7.

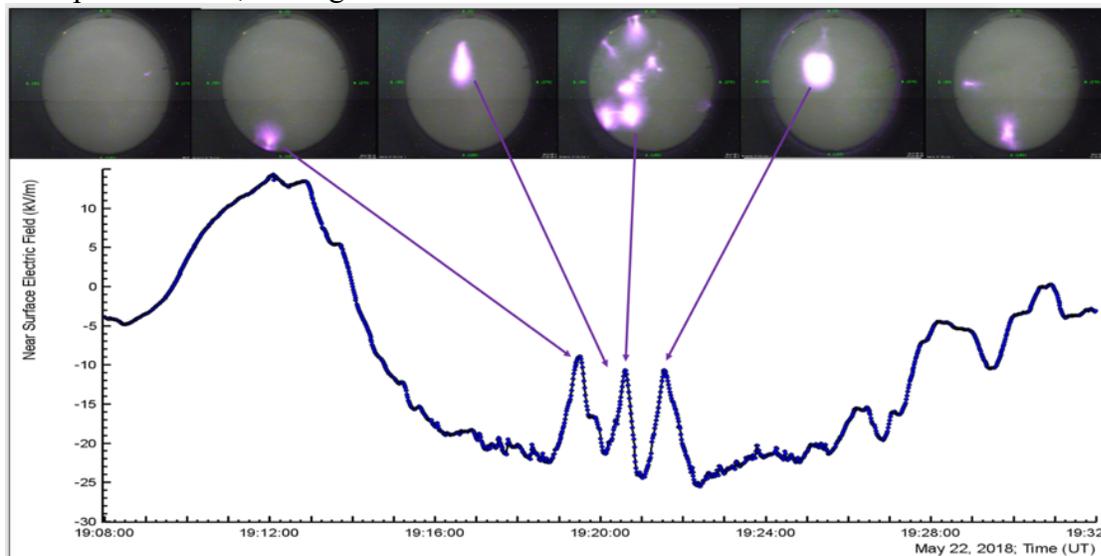

**Figure 6. The disturbances of the NSEF during TGE, measured by EFM-100 electric mill located on the roof of MAKET experimental hall (blue curve). In the upper panel, we show**

the 1-minute time series of the panoramic camera shots of the sky above the station; by a violet arrow, we indicate the times when shots were done.

In the upper panel, we show the shots of the panoramic camera, with intense violet lights that occurred during TGE. The light glows coincide with the NSEF outbursts and with the maximum of the TGE particle flux.

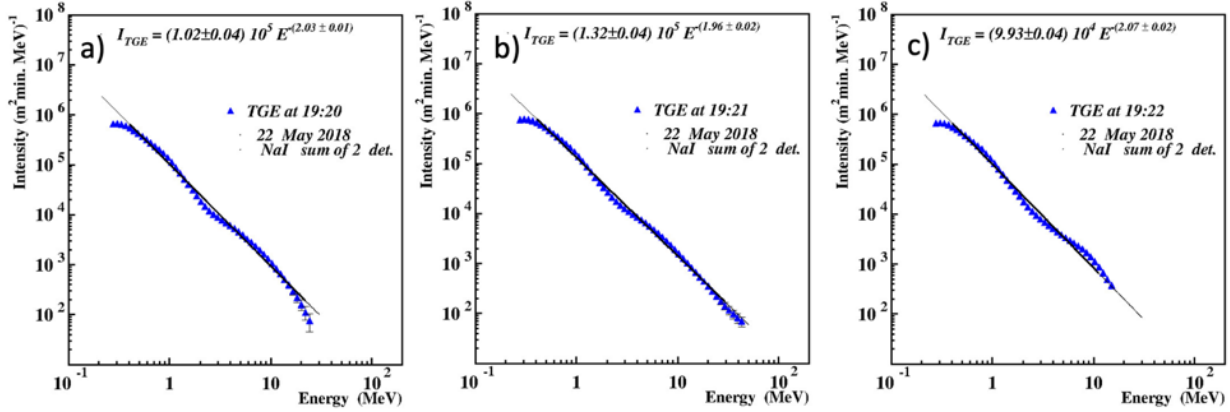

Figure 7. Differential energy spectra of TGE particles registered by NaI spectrometers at minutes of maximum flux.

In Fig. 7 we show the differential energy spectra measured by the NaI spectrometers during minutes of the maximum flux. 7 NaI spectrometers are located under the roof of the MAKET experimental hall (0.7 mm of the iron tilts) and 24/7 are monitoring the particle flux [26]. Large sizes (12 x 12 x 24cm) and low energy threshold allow measurements of the energy spectra from 0.3 MeV up to 100 MeV. At the lower energies (less than 2 MeV) we can see the contribution of the gamma radiation of $^{220}$Radon progenies lifted to the atmosphere by NSEF [27]. The energy spectra above 3 MeV are due to TGE particles, electrons, and gamma rays from the RREA avalanche reaching the earth's surface. The maximum energy of the TGE particles, reaching 50 MeV was measured at minutes 19:19 – 19:21 (Fig 7b), just in the middle of the disturbances of the NSEF and when most of the light glows occurred (upper panel of Fig. 6). Note that glows in the upper panel of Fig. 6 coincide with the "outbursts" of NSEF, which lower the net negative charge above the ground. Thus, after each strong discharge, the NSEF is impulsively enhanced and then returns to the previous value. And so on 3 times. 1-s clips of light glows above Aragats that occurred on 22 May 2018 can be found following the link [28].

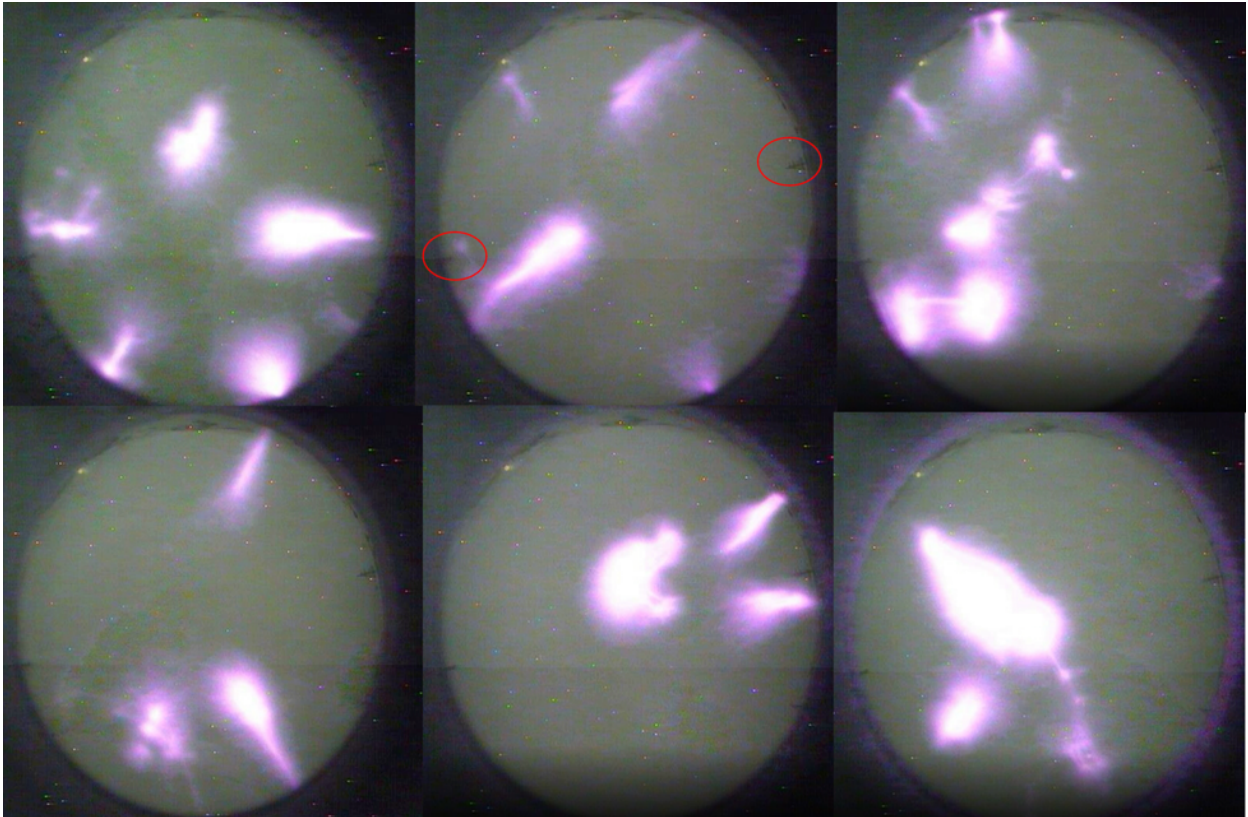

**Figure 8. Selected light glows occurred during maximum minutes of TGE (19:19-19:21), the full collection available from the 1-s clips [28]. By red ovals, we denote the 2 highest metallic structures on the station: the mast behind the MAKET experimental hall and the chimney behind the hotel building.**

In Fig. 8 we show light glows registered during the maximum of TGE flux only (there are many lights before and after TGE as well, see the links in the last column of the Table in the Mendeley dataset [28]). Possibly, the large particle fluxes and strong electric fields initiate discharges on high metallic masts around the Aragats station. However, we do not see in Fig. 8 any discharge on these structures (surrounded by red ovals). The ASC-N1 ALL SKY CAM model produced by Moonglow Technologies is a circular fisheye system providing 190 degrees hemispherical field of view (FOW).

The image sensor is a Color 1/3" Sony Super HAD CCD II with an effective pixel number across FOW is 546 x 457, with automatic exposure time (from $10^{-5}$ to 4s). The camera is located on the mast above the roof of MAKET experimental hall (8 m height above the ground), thus only the highest masts enter the camera's field of view (FoW). The lights located at the edge of the FoW are connected to the earth's surface, however, there are many shots appearing in the center of FoW. The glows appeared in different locations in FoW, thus, the discharges are appearing randomly in large regions of the sky below the clouds (cloud base height was ≈100 m, see left inset in Fig. 9). In Fig. 9, we show the camera location on the roof of MAKET building and nearby metallic constructions, that can be sources of corona discharges. However, after installing additional cameras at a distance of 120 m from the first one, we see that the same glow (shifted from each other according to camera location) appears in all 3 cameras, proving, that it is not a local corona discharge, but the sky glow. Additionally, we install a "spy" camera observing the vicinity of the panoramic camera and it did not observe any corona discharge.

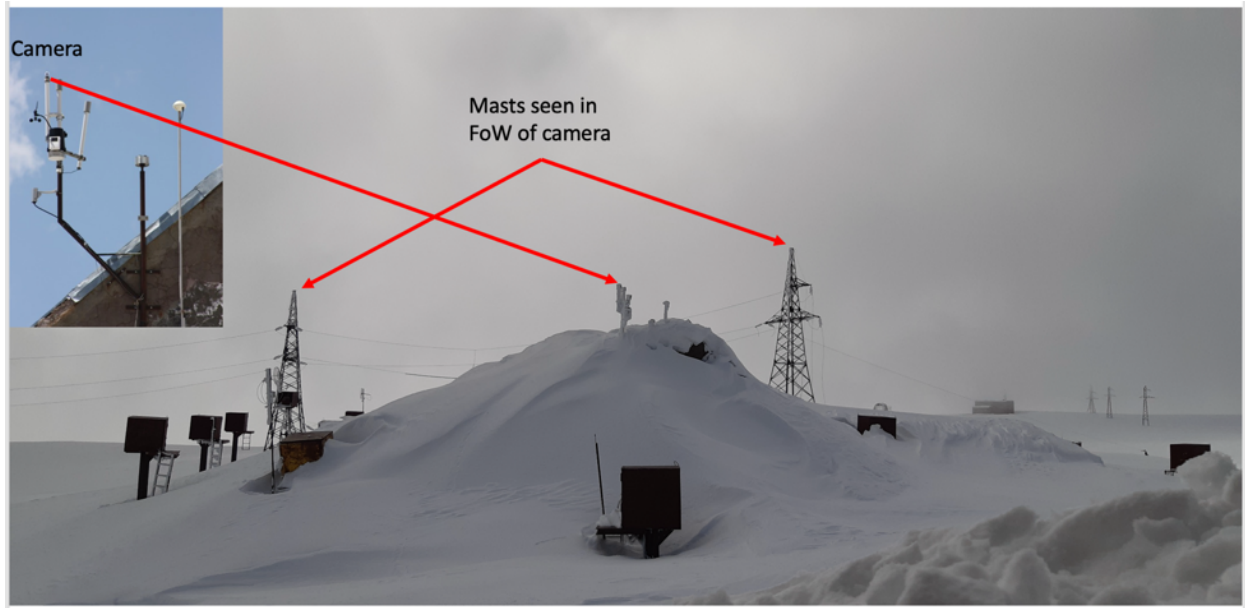

**Figure 9. The MAKET experimental hall in winter, in the inset the zoomed ALL SKY CAM surrounded by DAVIS weather station, BOLTEK's electric field sensor, and lightning tracker.**

In the following pictures, we show another very similar storm that occurred 3 days later; in the Mendeley dataset we post exhausting information on more than 10 similar storms, accompanied with light glows, the summary histograms will be shown in the discussion section.

As we can see in Fig. 10a rather a compact thunderstorm that occurred on 25 May 2018 also was approaching from the southwest. According to the WRF modeling results dense clouds were in the southwest of the Aragats station (Fig. 10b), in agreement with lightning data. Thus, there was an offset of the cloud cell's center from the Aragats station. However, as we can see in Fig. 10a, there were several lightning flashes nearby the station, and the radar reflectivity was as well modest, but non zero at the TGE time (Fig.10b). It is why we detect the graupel fall and LPCR, see Fig. 12. That is why the TGE wasn't terminated by a lightning flash and lasted for 14 minutes (fig. 11e); usually, TGEs are abruptly terminated by the nearby lightning flash [25].

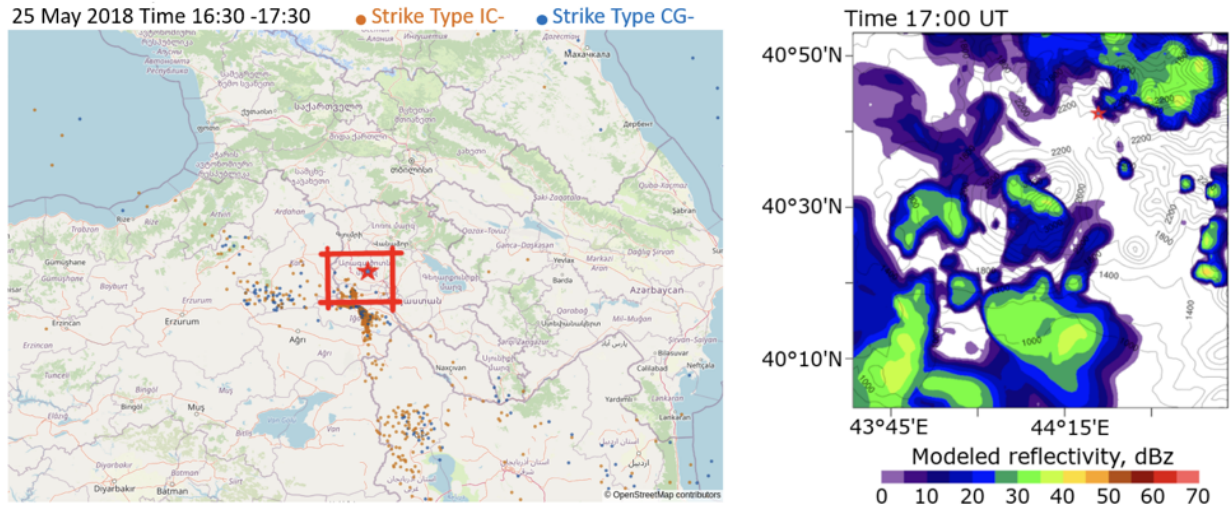

**Figure 10. Left: Projection of the thundercloud obtained by registering the lightning active zone with Boltek's storm tracker. Right: The maximum of the modeled with WRF radar reflectivity in the air column. The red star marks the location of the Aragats station.**

In Fig. 11a we show 1-sec time series of 3 units of STAND1 network (stand-alone 3-cm thick scintillators, see Fig.2); all 3 count rates of STAND1 network precisely correlate. It is interesting to see in Fig. 11b both the graupel (special spec on the glass of the panoramic camera) and light spots occurred at the maximum of the TGE flux. The specific specks on the shots are connected with graupel fall, as we demonstrate in the previous paper, see Figs 11 and 12 in [13]. In Fig. 11c we show 1-minute time series of the 3-cm thick scintillator belonging to the STAND3 detector (4 stacked scintillators located in the SKL hall) along with disturbances of NSEF and remote lightning occurrences.

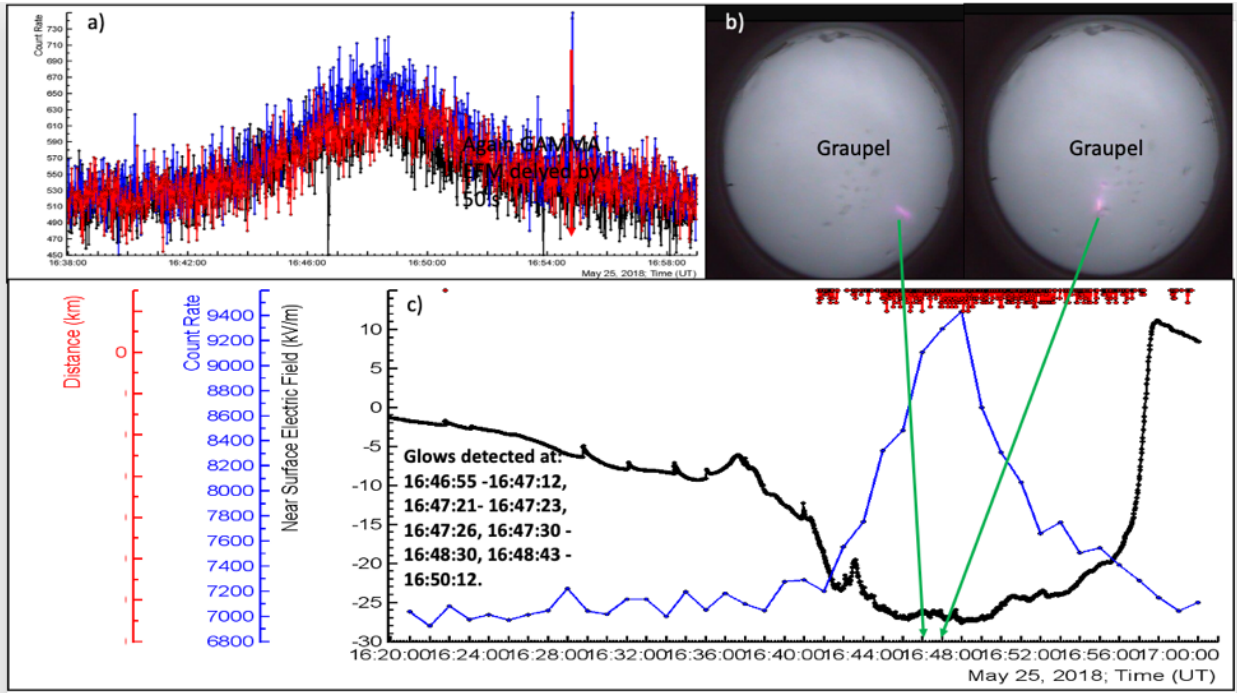

**Figure 11.** a) Time series of 1-sec count rates of 3 cm thick plastic scintillators of STNAD1 network, MAKET (black), GAMMa (blue), and SKL (red); b) panoramic camera shots with light spots and graupel fall observed at the maximum flux of TGE; c) disturbances of the NSEF measured by EFM-100 sensor located nearby GAMMA surface array, 1-minute time series of count rates of 3-sm thick plastic scintillator of STAND detector (indoor), red lines in the upper part of the frame – remote lightning flashes. In the inset we show the times of glows, by green arrows, the time of 2 glows which are shown in frame b).

In Fig. 12 we show WRF modeling results for the snow cluster (Fig .12a, and the graupel cluster Fig. 12b.) On 25 May, in contrast with May 22, the density of the graupel cluster was significant; the graupel hydrometeors were located very low above the station. We assume that the modeled graupel cluster likely formed an LPCR, which with the snow cluster (the main-negative layer – MN) forms a lower dipole accelerating electrons in the direction of the earth's surface. As we can see in Fig. 12 both clusters were in the same z-coordinates just above the station at the minutes of the maximum TGE flux and at the seconds when the light glows appear. Although the dipole MN-LPCR possibly plays a role in enhancing the electric field strength, the charge of the graupel cluster was quite small (because the NSEF is deep negative), thus, we see that the predominant impact is caused by the negatively charged snow cluster. The graupel cluster is much closer to the surface than the upper snow cluster (about 10 times closer, which leads to a 100 times higher field for the same charge), and still, it does not create a field comparable with field of the upper cluster. Therefore, we assume that the density of the graupel cluster is tens of times smaller than the density of the snow cluster.

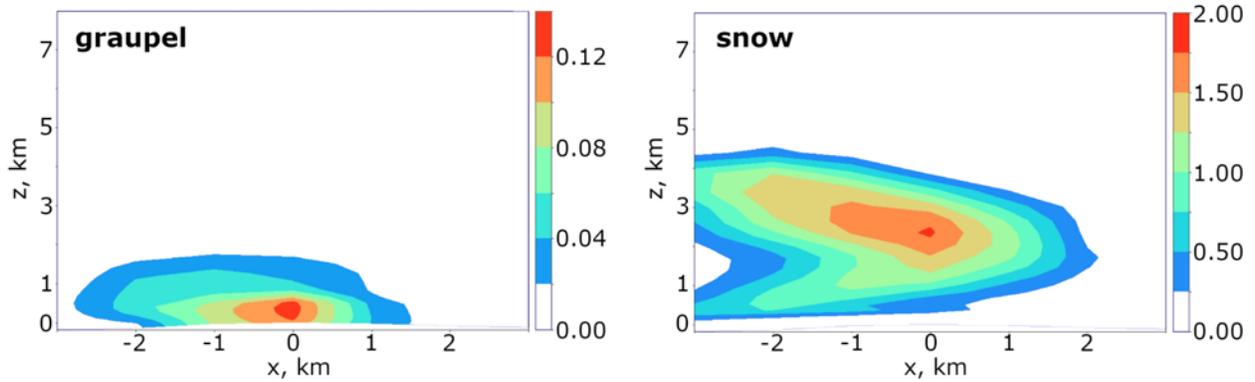

**Figure 12. The modeled spatial distribution of the graupel and snow particles at 16:50 UT on 25 May 2018. The vertical axes started on Aragats station height (3200 m), x-axes is oriented from west to east, the coordinate origin (zero) corresponds to the location of the Aragats station. The coloring characterizes the density of hydrometeors in ng/cm$^3$.**

In Fig.13 we show the 1-s time series of the NSEF, measured by MAKET electric mill. During TGE the NSEF was in the negative domain. In the upper panel, we show the shots of the panoramic camera that continuously monitors the skies above the research station. The graupel fall indicates that a strong electric field can be extended below cloud base due to the falling positively charged graupels that are in charge of the electric field production. In contrast with the previous storm, which occurred on 22 May, the light glows are not so bright (see Fig. 6), but considerably blurry. The height of the glows that are seen by panoramic cameras located on the ground, is restricted by the cloud base height that scattered and absorbed the optical emissions. We think that lights are seen via a cloud that on 25 May was extremely low (25-60 m); on 22 May cloud base height was estimated to be ≈90 m. and lights were very bright, thus, we can estimate very roughly the height of the light-glows to be 100 m.

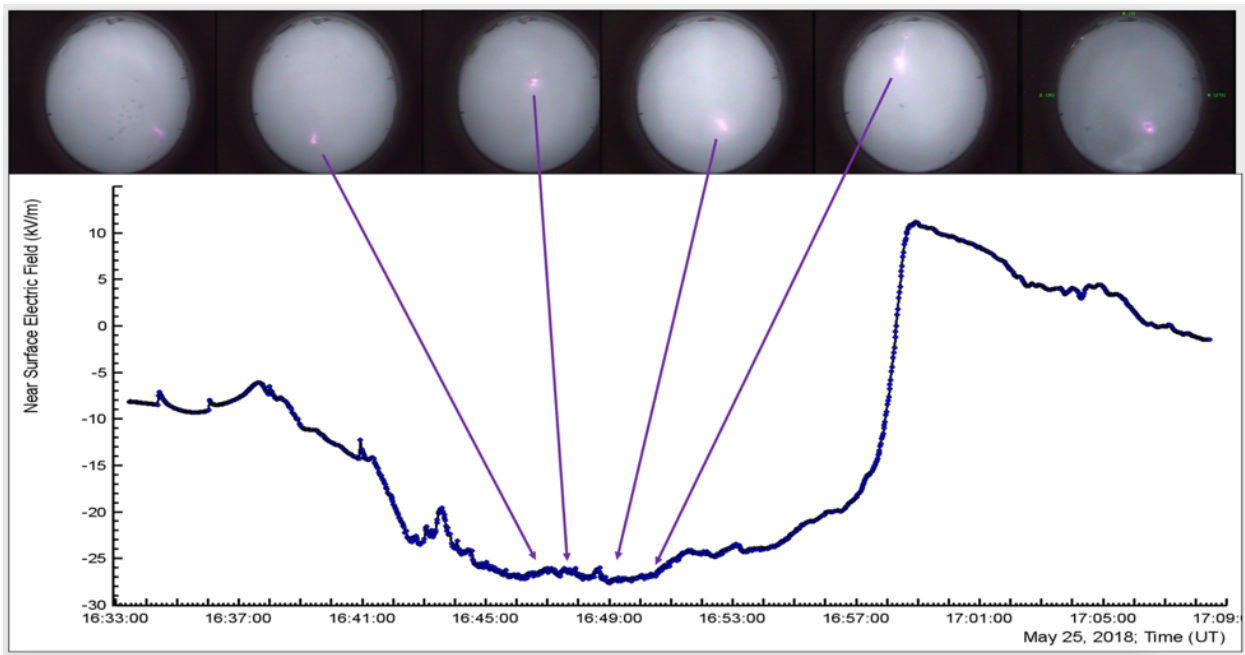

**Figure 13. The disturbances of the NSEF during TGE, measured by EFM-100 electric mills located on the roof of MAKET experimental hall (black), on the 13 m high mast (blue), above iron box hosted GAMMA detector scintillators (red), and above SKL experimental hall (green) In the upper panel we show the panoramic camera shots of the sky above the station; by violet arrows, we indicate the times when shots were done.**

4. Correlation analysis of the light glows, NSEF disturbances, and particle fluxes.

In this section, we will present the correlations of the light glows and NSEF in more detail to understand its possible origin. In Fig.14 we present the whole duration of a strong storm that occurred on 22 May 2018. The storm lasted nearly 4 hours, with multiple lightning flashes, it passes aside from Aragats station (see Fig.3), however, initiates 8 TGEs, 2 of which were significantly large (2 largest peaks in Fig 14). All 8 TGEs occurred when NSEF was in a deep negative domain approaching -20 kV/m. In previous sections, we use for the analysis of the data only one NSEF sensor located on the roof of the MAKET experimental hall, in this section, we will use data from all 4 electric field sensors to clarify the local effects of the charge rearrangements during TGE.

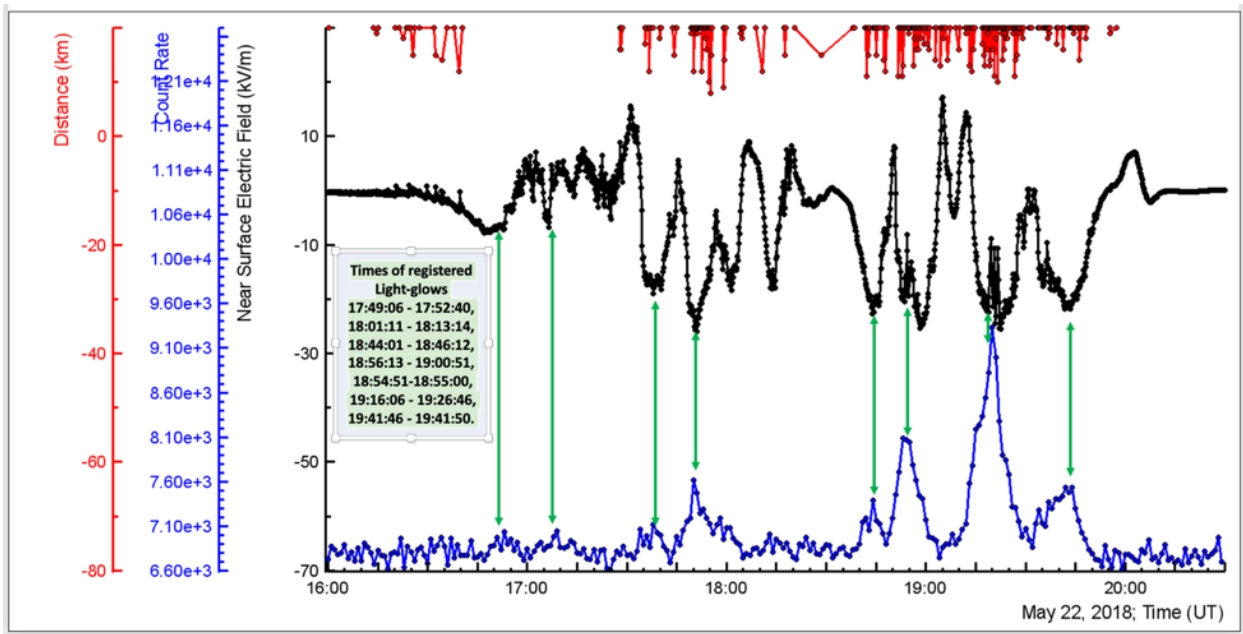

**Figure 14. Disturbances of NSEF – black curve, 1-min count rate of the STAND3 particle detector (energy threshold 4 MeV), blue curve, distance to the lightning flash – red lines. In inset – times of the light-glows.**

In Fig. 15 we show the disturbances of NSEF measured by all 4 sensors of the network. We can see periodic and symmetric progress of the NSEF during the TGE. 3 of the sensors located nearby on the highland show 3 peaks whose maximums precisely coincide in time. The remote sensor, located ≈300 m apart and ≈10 m lower in a valley exhibits 2 large peaks ahead of the rest 3 and didn't register the third.

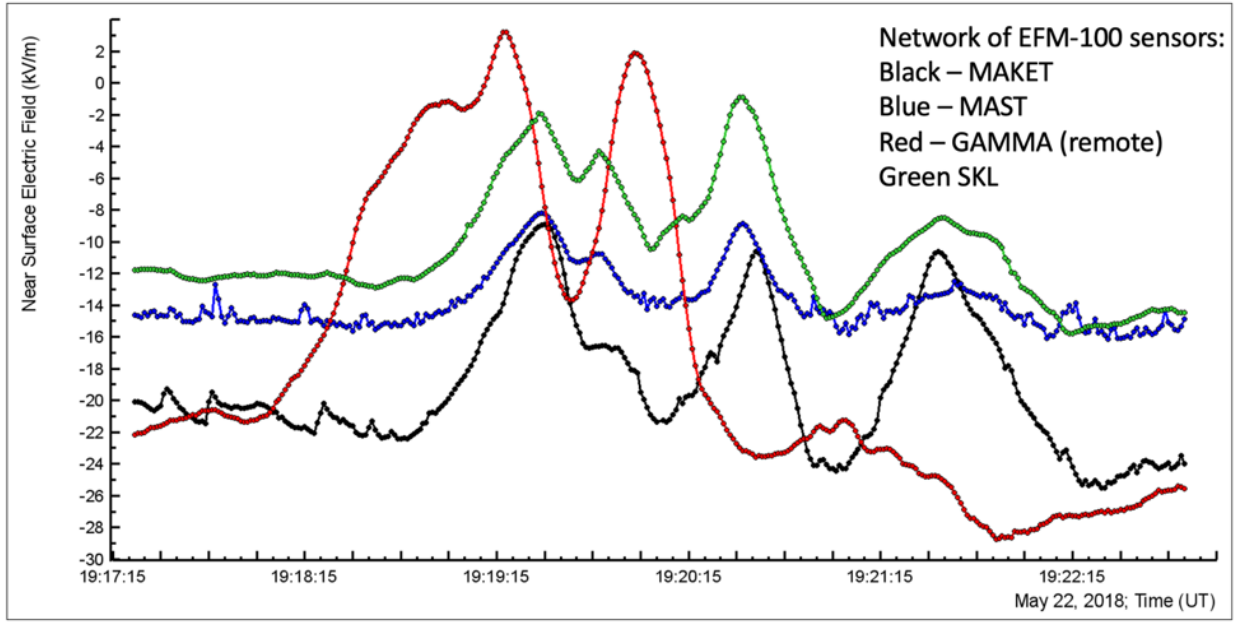

**Figure 15. Disturbances of NSEF during 6 minutes of TGE registered by 4 electric field sensors of EFM-100 type.**

In Fig. 16 we present a detailed comparison of measurements made by 2 nearby sensors, one on the MAKET experimental hall roof (black), and the second – on a standing alone 13-m high mast (blue). In the upper panel, we show the scatter plot and delayed correlations plot presenting the precise coherence of both measurements (the bias of delayed correlation is 0). In frames c-d, we show light glows, and in the frame 16g we outline the times of light glows by the lines of different colors. Thus, we can relate the periodic changes of the NSEF with the appearance of light glows in the FoV of the panoramic camera. The difference in amplitudes of NSEF changes possibly can be explained by the different locations of sensors. The sensor on the 13-m high mast possibly isn't enough sensitive to the charge rearrangement aloft, as the sensor, which is located 2 m above the metallic roof of the MAKET building.

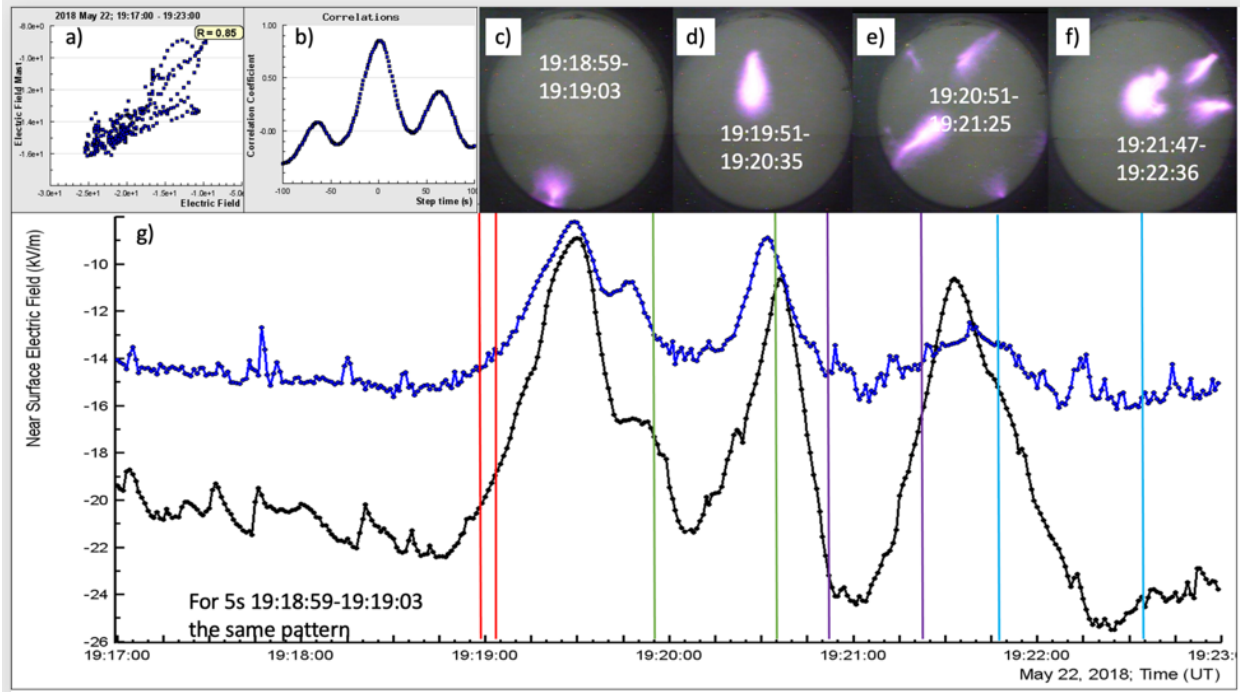

**Figure 16.** 1-s time series of NSEF disturbances measured by sensors located on the roof of MAKET building (black) and on standing alone 13-m heigh mast (blue). In the upper panel: a) scatter plot of 2 measurements, b) the delayed correlations plot, c-f - the typical patterns of light glows. By the colored lines are denoted the times of registered glows correspondent to patterns shown above.

In Fig. 17 we show the results of the correlation analysis of 2 remote NSEF sensors. From the delayed correlation plot we can see that the maximum of the electric field disturbances measured by the GAMMA EFM-100 located in a valley ≈ 300 m apart is 30 seconds ahead of the MAKET EFM (inset a). The amplitude of field disturbances is also larger for GAMMA EFM. Thus, we can conclude that field rearrangement first influences the GAMMA sensor, and only after 30 seconds – others.

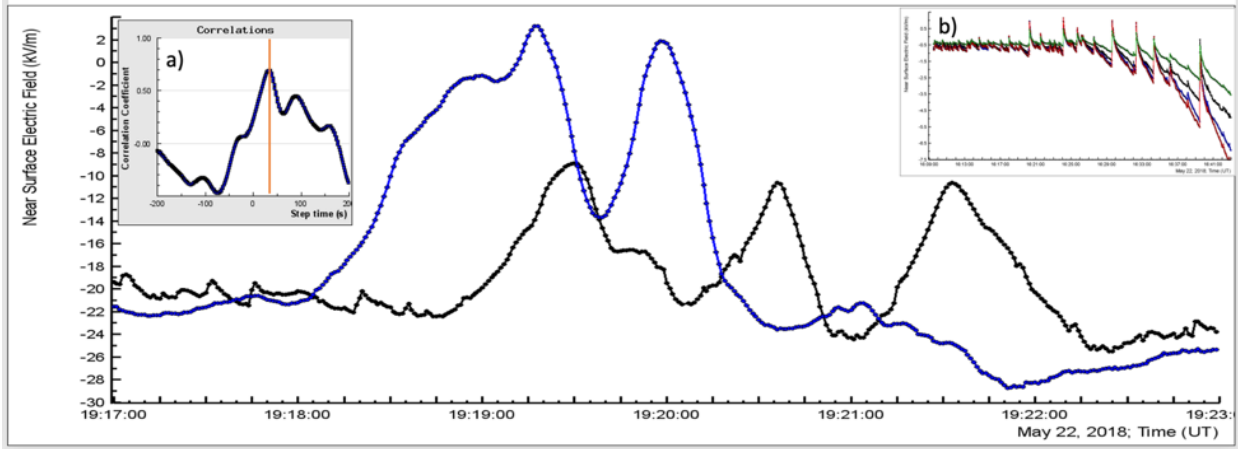

**Figure 17.** 1-s time series of NSEF disturbances measured by sensors located on the roof of MAKET building (black) and EFM-100 located above housing of GAMMA scintillators on

**1 m high mast. In inset a) – the delayed correlations plot, in inset b) – the typical pattern of NSEF disturbances to remote lightning flashes, registered at the beginning of the storm.**

The symmetric smooth shapes of analyzed NSEF disturbances are very different from the disturbances that originated from nearby lightning flashes. The nearby lightning discharges initiate the NSEF abrupt changes with the rise time of several hundreds of milliseconds and a long recovering tail of tens of seconds. Highly symmetric and many seconds long periodic structures surely cannot be related to any of the lightning discharge. The disturbances shown in Fig. 17 are not an exclusive case, in Fig. 18 we present another symmetric structure registered during a significant TGE that occurred a few minutes before the analyzed one. Again, the maximum "outburst" field measured by the GAMMA sensor is the same 30 seconds ahead of the MAKET sensor. However, this time amplitudes of both are equal.

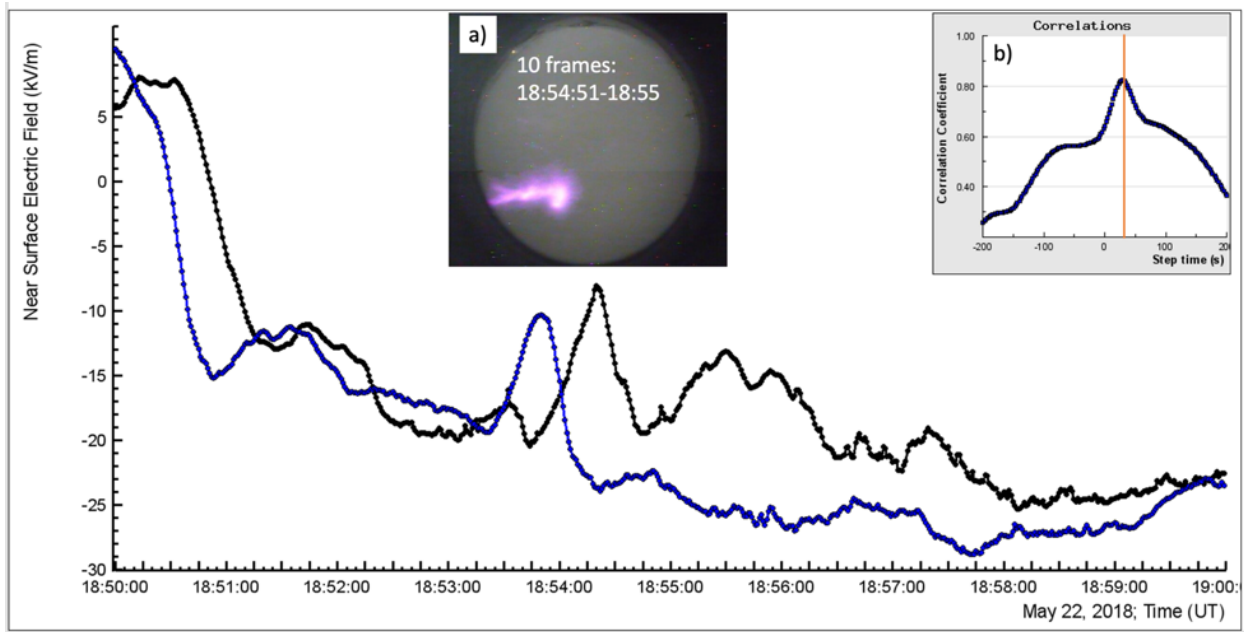

**Figure 18. 1-s time series of NSEF disturbances measured by sensors located on the roof of MAKET building (black) and EFM-100 located above housing of GAMMA scintillators on 1 m high mast. In the inset a) – the typic pattern of the light-glow measured during 10 seconds, b) – the delayed correlations plot.**

**Discussion and conclusions**

In our previous papers, we mentioned that TGEs are precursors of lightning flashes [29]. Recently we put a collection of 165 TGEs in the Mendeley datasets all terminated by the lightning flashes [25]. From this collection, supplied with links to explanatory materials, one can see that the nearby flashes (distances less than 3 km) terminate the TGE just at the beginning of the rising phase; the middle-distance flashes (3-8 km) - at the maximum or on the TGE decaying phase.

2 TGEs described in this paper are very different from TGEs terminated by lightning flash. Centers of the storm lightning activity were ≈10 km apart from the detectors and lightning flashes do not catastrophically lower the potential difference and abruptly terminate the RREA process in the thunderous atmosphere above the particle detectors. Instead, numerous weak discharges do not terminate particle fluxes that continued for 14 and 18 minutes and originate light glows during maximum flux of TGE. These considerably weak electrical discharges do not trigger the Aragats system of electromagnetic pulse detection, which is triggered only by strong nearby lightning flashes. From our analysis, we deduce that the "outburst" of NSEF first was measured by the remote EFM sensor located ≈300 m apart from others and 10 m lower in the lowland open to the Ararat valley. The 30 seconds delay possibly indicated some local charge rearrangement that is due to the large negative electric field.

Lightning active zones miss the station for both TGEs as is seen in the lightning location and WRF maps (Figs 3 and 10), and nearby lightning flashes occurred at distances of ≈10 km, do not terminate TGEs. The particle flux was stable at hundred meters scale, as we can see from precisely correlated particle flux enhancements measured by remote detectors (see Figs 4 and 11). It indicates that the intracloud electric field that originates particle fluxes also was stable on the minutes time scale.

Thus, if the storm is just above particle detectors nearby lightning flashes terminate RREA after a few minutes [25]. When the storm active zone is far away from particle detectors (>10 km) the TGE extends for tens of minutes and smoothly terminates when conditions of the atmospheric electric field fail to support RREA. In Fig 19a we show the distribution of distances to lightning flashes abruptly terminated by the lightning flash [25]. In Fig. 14b we show the same distribution for long-lasting TGEs, accompanied by light glows (a collection of these TGEs is posted in [28]). It is apparent that these distributions are belonging to the distinct classes; only the second class (corresponding to the distribution presented in Fig.19b) supports the origination of light glows.

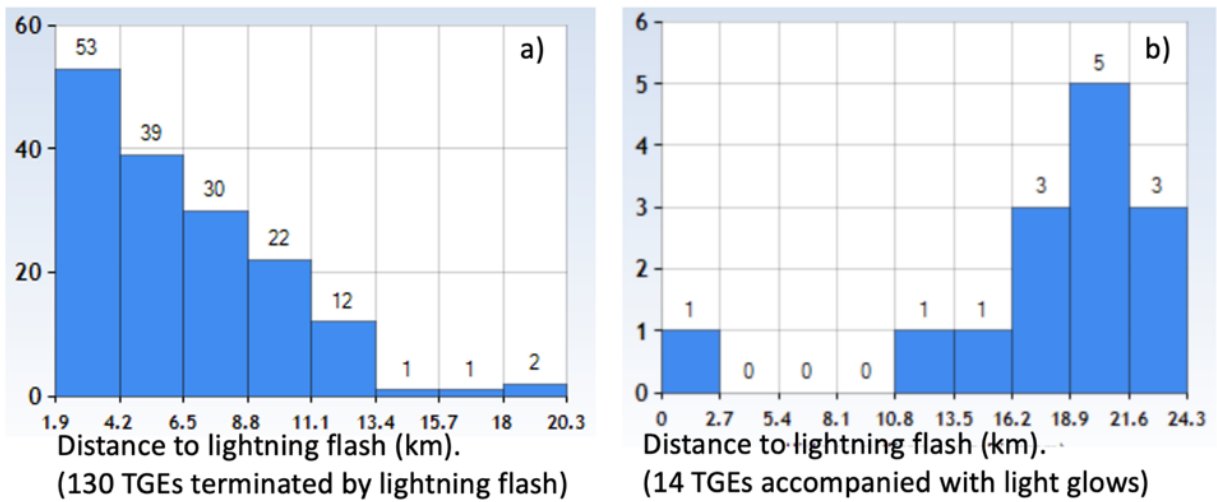

Figure 19. Distribution of the distances to the lightning flash for TGEs terminated by a flash a), and for TGEs accompanied by the light glows b).

Thus, the electron acceleration occurred in both cases and RREA particles cover large areas below thunderclouds. We conclude, that the RREA can be unleashed in a very large spatial domain around the active lightning zone and many kilometers apart. From the energy spectra of a TGE registered on May 30, 2018 [30] we estimate the total number of gamma rays (with energies above 300 keV) hitting the earth's surface to be $1.3*10^6/m^2$min. Assuming that ≈ 2000 thunderstorms are active on the globe and that the overall surface of the thunderous atmosphere each moment can be estimated as 2.000 * 100 km² = 200,000 km² (0.04% of the globe surface), we come to an estimate of $≈10^{16}$ gamma rays are hitting the earth's surface each second! Observed light glows are not local corona discharges on the camera mast but a discharge in the atmosphere above the Aragats station influencing all electric sensors. These discharges do not initiate lightning flash, only large disturbances of the NSEF and, light glows in the sky above the station. During the most of "glow" events, the NSEF was in the deep negative domain, only 3 from 14 were in the positive domain, see Fig. 20. The origin of light glows is under discussion, the possible explanations are intense fluxes of TGE electrons [31, 32], ball lightning [33,34], St. Elmo's fires, and geomagnetic disturbances [35]. However, after examining luminous TGE events, along with lightning location maps and NSEF time series, we think that these unusual luminous phenomena below thunderclouds are a new optical phenomenon. An electrical discharge much weaker than a lightning flash could only partially neutralize the charge above, and hence, only partially lowers the corresponding potential difference, allowing the electron accelerator to operate and send particle fluxes in the direction to the earth's surface. Simultaneously, these types of discharges initiate light glows in the thunderous atmosphere inside and below thunderclouds.

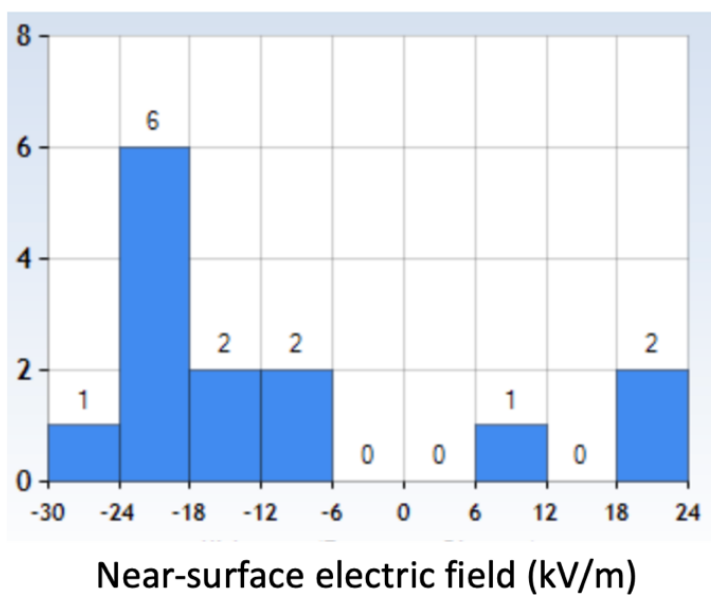

**Figure 20. The NSEF distribution for the TGE events accompanied by light glows**

**Acknowledgments**

We thank the staff of the Aragats Space Environmental Center for the uninterruptible operation of experimental facilities on Aragats under severe weather conditions. The data for this study is available in numerical and graphical formats by the multivariate visualization software platform ADEI on the WEB page of the Cosmic Ray Division (CRD) of the Yerevan Physics Institute, http://adei.crd.yerphi.am/adei and from Mendeley datasets. The authors acknowledge the support of the Science Committee of the Republic of Armenia (research project No 21AG- 1C012), in the modernization of the technical infrastructure of high-altitude stations. We also acknowledge the support of the Basic Research Program at HSE University, RF.

**Declaration of Competing Interest**

The authors declare no conflict of interest.

**Data Availability Statement**

The data for this study are available in numerical and graphical formats on the WEB page of the Cosmic Ray Division (CRD) of the Yerevan Physics Institute, http://adei.crd.yerphi.am/adei and from Mendeley datasets [14,26].